\documentclass[]{aastex631}

\usepackage{url}
\usepackage{verbatim}
\usepackage{natbib}
\usepackage{amsmath}

\shorttitle{Eldest X-ray SNe}
\shortauthors{Ramakrishnan and Dwarkadas}

\begin{document}


\newpage

\title{X-ray Luminosity of Decades-Old Supernovae}

\correspondingauthor{Vikram V. Dwarkadas}
\email{vikram@astro.uchicago.edu}

\author[0000-0002-9176-7252]{Vandana Ramakrishnan}
\affiliation{Department of Astronomy and Astrophysics, University of
  Chicago\\ 5640 S Ellis Ave, Chicago, IL 60637}
\affiliation{Current Address: Department of Physics and Astronomy, Purdue University \\
525 Northwestern Ave, West Lafayette, IN 47907}

\author[0000-0002-4661-7001]{Vikram V. Dwarkadas}
\affiliation{Department of Astronomy and Astrophysics, University of
  Chicago\\ 5640 S Ellis Ave, Chicago, IL 60637}

\begin{abstract}
    The transition from supernovae (SNe) to supernova remnants (SNRs) remains poorly understood, given the age gap between well-studied examples of the two. In order to bridge this gap, we analysed archival Chandra data for some of the oldest supernovae detected in X-rays, in order to extend their light curves out to late times. We fitted the spectra with thermal models.  All the SNe with multiple X-ray data points were found to have similar X-ray luminosity, which was decreasing with time. The X-ray luminosity will likely continue to decrease while the SNe are evolving in a wind medium, but is anticipated to increase in the Sedov phase when the SNe are interacting with a constant density interstellar medium, bringing it in line with observed SNRs
\end{abstract}

\section*{}

While both supernovae (SNe) and supernova remnants (SNRs) are well-studied objects, the transition from one to another is still poorly constrained. Theoretical models for the shock evolution exist, but are yet to be verified with observational evidence. This is in large part due to the fact that a considerable age gap exists between well-studied SNe and SNRs. The majority of SNe are only followed up observationally for months to years, while the youngest known Galactic SNR, G1.9+0.3 is $\sim$ 100 years old \citep{Reynolds2008}, and others are $>$ 300 years. The obvious remedy is to study decades-old SNe on their way to becoming a remnant. 

X-rays are ideal to study SN evolution. The thermal X-ray emission from SNe is caused by the shock-heating of the surrounding medium as the SN shock travels through it. Tracing the X-ray light curve can give information on the evolution of the shock velocity, electron temperature and the density of the surrounding medium. 

In this work we study Chandra archival observations of the oldest-known X-ray SNe, SN 1941C, SN 1957D, SN 1968D and SN 1970G. Chandra is ideal for this purpose because of its high spatial resolution and point source sensitivity, allowing limited counts from old SNe to be detected above the background (e.g. \citet{Soria_2008}). We build on past studies of these SNe \citep{Immler_2005, Soria_2008, Long_2012} to extend their light curves over longer time baselines. We calculate the 0.3-8 keV X-ray flux using the \textsc{ciao} software (v 4.10.0), fitting the spectra with \textsc{xspec} models where possible, and estimating the flux from the count rate where the counts are insufficient. In general, we fit source and background models simultaneously to the ungrouped spectrum using the Cash statistic. An exception is the archival 2004 observation of SN 1970G, where the large number of counts allowed us to subtract out the background and group the counts by 15. The background was modeled with either a powerlaw or polynomial model, depending on which model was visually confirmed to be a better fit. In all cases, a single thermal \textit{vmekal} model was used to fit the spectra. When the counts were not high enough for spectral fitting, the count rate, or the upper limit on the count rate, was extracted using the \textsc{ciao} command \textit{srcflux}, and extrapolated to estimate the flux using the online \textsc{pimms} tool. The parameters used to estimate the flux in this case were chosen based on the closest temporal observation where spectral fitting was possible. SN 1941C had only a single archival observation, but for the remaining three SNe we were able to construct the light curves out to later times than had been done in past studies.

{\bf Results:}. (1) Similar to the majority of observed SNe \citep{Dwarkadas_2012, vvd14, drrb16, Bochenek2018}, the light curves of SN 1957D, SN 1968D and SN 1970G are consistent with a steady powerlaw decrease with time, of the form L$_X$ $\propto$ t$^{-\alpha}$, albeit with large errorbars on the powerlaw index (Figure 1, left). We find the best fit values of the powerlaw index $\alpha$ to be 2.22 $\pm$ 1.35 for SN 1970G, 1.47 $\pm$ 2.81 for SN 1968D and 2.24 $\pm$ 1.56 for SN 1957D. In contrast to published results \citep{midlife_crisis}, we do not find any increase in the luminosity of SN 1970G in 2011. This is likely attributable to modeling differences. \citet{midlife_crisis} model the 2011 observation with a powerlaw, and compare it to past determinations of the flux \citep{Immler_2005} which used a thermal model. We model all observations of SN 1970G with a thermal model, since this fits the 2004 spectrum the best. If instead a powerlaw model is used to model all observations, it consistently yields higher fluxes compared to a thermal model at any given epoch. However modeling all observations with a powerlaw still results in a decreasing flux with time.  

(2) These SNe all have comparable luminosity, within two orders of magnitude (in contrast to the at least 10 orders of magnitude variation in the SN X-ray luminosity overall - see, e.g. Figure 1 of \citet{Dwarkadas_2012}). This is despite the fact that their electron temperatures vary by a factor of a few ($\sim$ 0.4 keV for SN 1970G, $\sim$ 1 keV for SN 1968D, $\sim$ 3.7 keV for SN 1957D and SN 1941C; see \citet{Ramakrishnan_2020} for details). 

(3) The luminosity of the SNe considered here is at present comparable with that of many Galactic Type II supernova remnants (SNRs). If the present trend of decreasing luminosity continues, then the luminosity of these SNe would be considerably \emph{lower} than that of the majority of core-collapse remnants in a few hundred years. It is possible that the observed SNRs arise from more X-ray luminous SNe, but if so the question arises as to why such luminous SNe have not been discovered while fainter ones have. An alternative is that the luminosity of these decades-old SNe increases in the Sedov-Taylor phase. This has been rigorously demonstrated by \citet{Hamilton1983}, but can be simply understood using a back-of-the-envelope calculation as follows.\\
The X-ray luminosity can be approximated by:
\begin{eqnarray}
    L_X = \Lambda n_e^2V
\label{eqn:xray}
\end{eqnarray}
Where $\Lambda$ is the cooling function, n$_e$ is the electron density and V is the emitting volume. In the Sedov-Taylor stage, for a shock expanding into a medium whose density evolves as $\rho$ $\propto$ $r^{-s}$, the self-similar solution for the shock radius $R_{ST}$ is given by the Sedov-Taylor solution,  
\begin{eqnarray}
    R_{ST} &\propto& t^{2/(5-s)}
    \nonumber \\
    \implies {\rho}_{ST} &\propto& t^{-2s/(5-s)}
\end{eqnarray}
The cooling function  $\Lambda$ below about 2.6 $\times$ 10$^7$ K (representative of remnants in the Sedov-Taylor phase) has the form $\Lambda$ $\propto$ $T_e^{-0.5}$ \citep{Chevalier2017}. For a strong shock, T$_e$ $\propto$ v$_{sh}^2$ and v$_{sh}$ $\propto$ r$_{sh}$/t. So, the X-ray luminosity evolves with time as,
\begin{eqnarray}
    L_X &\propto& t^{-4s/(5-s)}\left(\frac{r}{t}\right)^{-1}r^3
    \nonumber \\
    L_X &\propto& t^{(9-5s)/(5-s)}
\end{eqnarray}
As can be seen from equation 3, if a remnant is expanding into a medium whose density profile is flatter than s = 1.8, {\em the X-ray emission will increase with time}. In particular if the remnant is expanding into the constant density interstellar medium, i.e. s = 0, the luminosity L$_X$ $\propto$ t$^{1.8}$. Thus we would expect that the remnant luminosity will rise again in the Sedov-Taylor stage.\\

\begin{figure}
    \centering
    \includegraphics[width=\linewidth]{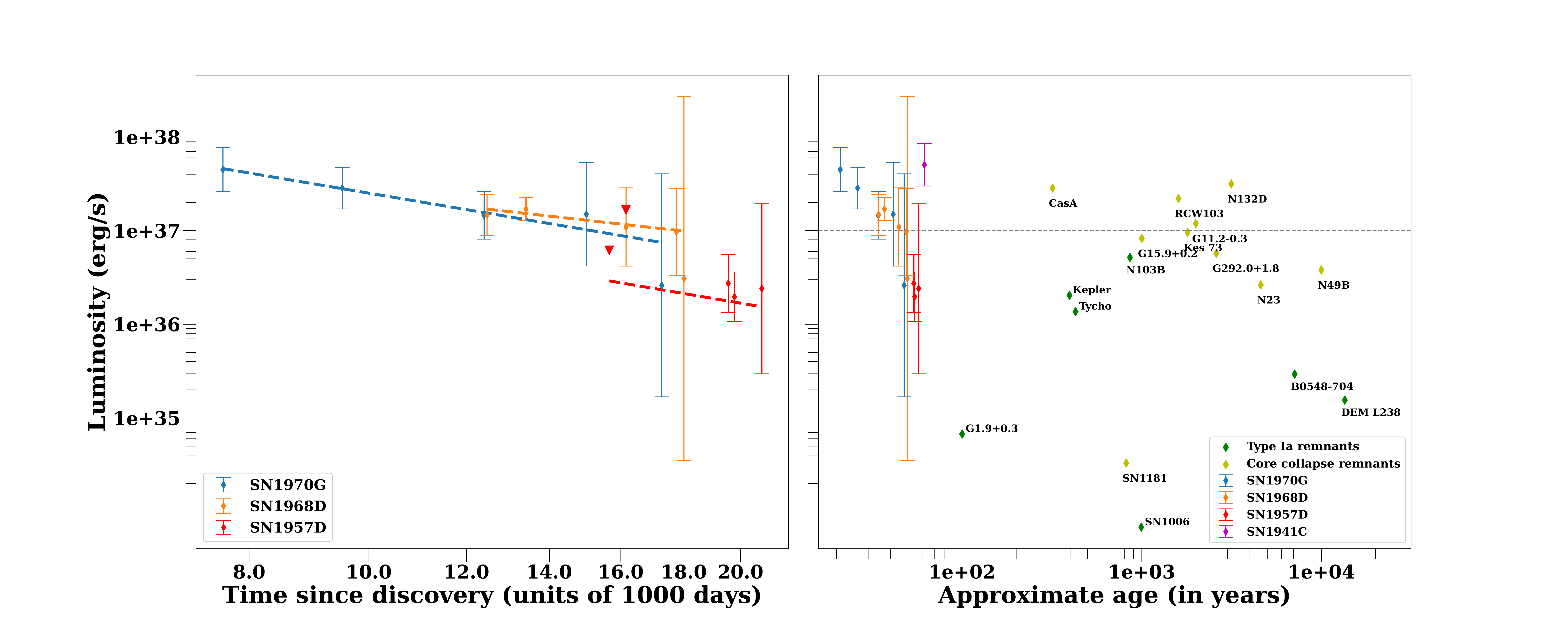}
    \caption{\textit{Left:} The 0.3 - 8 keV X-ray luminosities of SN 1957D, SN 1968D and SN 1970G, plotted as a function of time and showing the best-fit powerlaw model (dotted lines). \textit{Right}: The luminosity of the SNe studied in our work, plotted in comparison to some Galactic and LMC SNRs. The luminosity of these old SNe is presently comparable to that of the brightest core-collapse remnants, as indicated by the grey dotted line. }
\end{figure}

{\bf Acknowledgements}: This work was supported by NSF Award 1911061 to PI V. Dwarkadas at the University of Chicago.

\bibliographystyle{aasjournal}
\bibliography{main2}



\end{document}